\documentclass{elsart}
\usepackage[sumlimits]{amsmath}
\usepackage{latexsym}
\usepackage{amsfonts}
\usepackage{amssymb}
\usepackage{euscript}
\theoremstyle{definition}

\usepackage{bm}
\begin{document}
\tolerance=5000
\def\be{\begin{equation}}
\def\ee{\end{equation}}
\def\bea{\begin{eqnarray}}
\def\eea{\end{eqnarray}}
\begin{frontmatter}

\title{Towards the topological quantization of classical mechanics}

\author{Francisco Nettel, Hernando Quevedo and Moic\'es Rodr\'\i guez}
\address{   Instituto de Ciencias Nucleares,
     Universidad Nacional Aut\'onoma de M\'exico \\
     A. P.70-543, M\'exico D.F. 04510, M\'exico}

\date{\today}

\begin{abstract}
We consider the method of topological quantization for conservative systems with a finite number of degrees of freedom. Maupertuis' formalism for classical mechanics provides an appropriate scenario which permit us to adapt the method of topological quantization, originally formulated for gravitational field configurations. We show that any conservative system in classical mechanics can be associated with a principal fiber bundle. 
As an application of topological quantization we derive expressions for the topological spectra of some simple mechanical systems and show that they reproduce the 
discrete behavior of the corresponding canonical spectra. 
\end{abstract}
\end{frontmatter}

\section{Introduction}
\label{sec:int}
Inspired by Dirac's idea that leads to the quantization of the electric charge \cite{dirac} and the underlying geometric and topological aspects, Pati\~no and Quevedo introduced in \cite{lh,patquev} the method of topological quantization for gravitational fields. The main geometric concept in Dirac's work that leads to the quantization of the electric charge is the existence of a complex line bundle (the electromagnetic bundle) for an electron moving in the field of a magnetic monopole. The quantization condition arises from demanding that the connection is well defined everywhere on the bundle. The magnetic field has a singularity in the origin where the magnetic monopole is situated making the bundle non trivial. Therefore, the base space must be covered by (at least) two coordinate patches. In the overlap of two such patches the transition function must be single valued, leading to the quantization condition. In the case that more than two intersecting patches are considered, one must analyze the transition functions and fulfill their usual properties. It can be shown that this condition also follows from a general Gauss-Bonnet theorem for complex line bundles over $\mathbb{R}^3$ with the origin removed. The quantization condition 
is obtained by integrating  the Chern form of the bundle under examination, over a closed oriented surface embedded in the base space. 
For an interesting account on this subject see \cite{frankel}. 

Based on these general geometric concepts, in \cite{patquev} the method of topological quantization for gravitational fields was proposed, provided that the analyzed field configuration can be represented as a principal fiber bundle with a spin connection. It was shown there that any solution of Einstein's vacuum field equations can be represented by a unique principal fiber bundle with the spacetime as the base space, the structure group $SO(1,3)$ (isomorphic to the standard fiber), associated to the Lorentz invariance of the underlying orthonormal frame, and an one-form connection which takes values in the algebra of the group $SO(1,3)$. If there is a gauge matter field minimally coupled to the Einstein-Hilbert Lagrangian, 
an additional standard fiber can be associated to the group of gauge transformations, and an additional connection appears that takes values in the algebra of the gauge group.  Having this geometric structure as starting point, one can find a discrete relationship among the physical parameters which describe the field configuration. As in Dirac's quantization, this is done by demanding that the principal fiber bundle exists, that is, all its elements are well defined. In the case of a non trivial bundle, we are compelled to cover the base space with more than one open subset, and the conditions for being well defined are given by the properties of the transition functions. 
This can be used to construct a spin connection that is well defined throughout the principal fiber bundle. As we mentioned above in the example of Dirac's approach, the quantization condition can also be achieved by calculating 
the topological invariants of non trivial principal fiber bundles which may come out by removing points from the base space. This is the case, for instance, in gravitational fields with points of singular curvature. The discretization conditions we obtain by investigating either the topological invariants or, equivalently, the geometric structure of the principal fiber bundle is what we call the topological spectrum of the underlying classical configuration \cite{netquev07}.

The aim of this work is to adapt the method of topological quantization to classical conservative systems with a finite number of degrees of freedom. To accomplish this, we need a formulation of classical mechanics that permit us to construct a principal fiber bundle over a base manifold that contains all the information about the classical system. This construction will eventually lead to a quantization condition. In the most common variational approaches, that is, Hamilton's principle in the tangent bundle (the Lagrangian approach) and in the cotangent bundle (Hamiltonian approach), the formulation is over a geometric structure which is a vector bundle. However, these vector bundles are not unique, and they do not suit our purposes. We shall see that an alternative description due to Maupertuis is adequate for the construction of a principal fiber bundle. We present some preliminary results towards a complete scheme of quantization where we hope that, in a no so far future, states and dynamics will be incorporated in the formalism. 

This work is organized as follows. 
 In section \ref{sec:mau} we briefly review Maupertuis' formalism of classical mechanics.  
In section \ref{sec:geo} we analyze alternatives for the construction of the principal fiber bundle for classical mechanics and we discuss why Maupertuis' approach is an option which fits our purposes. Then we examine the base space from a geometric point of view and we cast the relevant equations using Cartan's formalism. In section \ref{sec:pfb} we state a theorem that assures the existence and uniqueness of the principal fiber bundle which represents a conservative mechanical system. Section \ref{sec:hosc} is devoted to define the topological spectrum of a classical mechanical system. 
Some examples are given which illustrate this definition. 
Finally, in section \ref{sec:con} we present some concluding remarks and we address a few questions for future work.

\section{Maupertuis formalism in classical mechanics}
\label{sec:mau}
In classical mechanics the equations of motion and a set of initial conditions constitute 
all the information that is needed to describe a physical system. Of course, once we have
solved the equations of motion we truly say that we know completely the behavior of that system. 
Variational principles in this context are known as Hamilton's principle and, except for pathological cases (e.g. dissipative systems or nonholonomic constraints), give an alternative way for finding the equations of motion. In this sense we say that they are equivalent to the Newtonian formulation. 
Consider the configuration space $M$ with 
coordinates $q^\alpha$, $\alpha=1,2,...,k$. 
Hamilton's principle can be formulated either using the Lagrangian function $L=L(q,\dot{q},t)$ on the tangent bundle $TM$, or from an 
one-form $p_{\alpha}dq^{\alpha}$ on the cotangent bundle (phase space) $T^*M$  and the Hamiltonian function $H=H(q,p,t)$. 
These two formulations are equivalent as long as the canonical momenta, $p_{\alpha} = \frac{\partial L}{\partial \dot{q}^\alpha}$,  are invertible, 
and we say that they are related by a Legendre transformation.

Consider the extended contangent bundle $T^*M \times \mathbb{R}$ with coordinates $(q^\alpha, p_\alpha, t)$ and the Hamiltonian function $H=H(q,p,t)$. 
Then, Poincare's one-form can be defined as $p_\alpha dq^\alpha - Hdt$. 
Suppose there is a curve $\lambda = (q(t), p(t))$ on the phase space with fixed endpoints, which represents the physical path of the system. Hamilton's principle in phase space can be stated as follows. For all possible curves with fixed endpoints $\delta q = 0$ and $\delta t = 0$, the integral $\int (p_\alpha dq^\alpha - Hdt)$ has an extremum in $\lambda$, 
\be
\label{hamps}
\delta \int (p_\alpha dq^\alpha - Hdt) = 0\ . 
\ee
This leads to Hamilton's equations of motion. When the Hamiltonian does not depend explicitly on time, it defines a hypersurface on the phase space $H(q,p)= E$. In this case, from the equation above we can see that the variational principle can be formulated for the action $\int p_\alpha dq^\alpha$ for curves lying on this hypersurface. The phase flow, that is, the curve $\lambda = (q(t), p(t))$ which is solution of Hamilton's equations, with initial conditions on the hypersurface $H=E$, will remain on that hypersurface. Projecting $\lambda$ on the configuration space $M$, we obtain a curve $q^\alpha$ which traces out the physical path on $M$. 
Now we can recast the variational principle as follows. Under variations of the curve $q^\alpha = q^\alpha(t)$ with fixed endpoints $\delta q = 0$, parametrized in such a way that $H=E$, the reduced action integral $\int p_\alpha dq^\alpha$ is an extremal  of the physical trajectory, 
\begin{equation}   \label{mauprinc}
\delta \int p_\alpha dq^\alpha = 0.
\end{equation}
This is Maupertuis' principle of least action. Consider the Lagrangian 
\be
L = \frac{1}{2} g_{\alpha\beta} \dot q ^\alpha \dot q ^\beta - V(q) \ ,
\label{lag}
\ee
where $g_{\alpha\beta}=g_{\alpha\beta}(q)$ is a Riemannian metric on 
the configuration space. The corresponding Euler-Lagrange equations lead to the 
geodesic-like equations  
\be
\ddot q ^\alpha + \Gamma^\alpha_{\ \beta\gamma} \dot q ^\beta \dot q ^\gamma 
+ g^{\alpha\beta} \partial_\beta V(q) = 0\ ,
\label{geo1}
\ee
where $\Gamma^\alpha_{\ \beta\gamma}$ is the Levi-Civita connection associated to the metric $g_{\alpha\beta}$.
The canonical momenta are $p_\alpha = g_{\alpha \beta}\dot{q}^\beta$, and since $H=E$, we have that $T=E-V(q)$. Then, the ``reduced'' Poincar\'e's one-form is
\begin{equation}   \label{redpoinform}
p_\alpha dq^\alpha = \frac{\partial L}{\partial \dot{q}^\alpha}\dot{q}^\alpha dt = 2 T dt = 2[E-V(q)]dt \ .
\end{equation}       
On the other hand, the kinetic energy can also be defined in terms of the arc length 
of a curve in the configuration manifold $M$, 
\begin{equation} \label{arclent}
T = \frac{1}{2}\left(\frac{ds}{dt}\right)^2 \ ,
\end{equation}
so that the integrand of the reduced action can be written as $
p_\alpha dq^\alpha = 2Tdt = \sqrt{2T}ds$, 
where $ds^2$ is the line element in  the configuration manifold $M$. 
Furthermore, we introduce the Jacobi metric as
\begin{equation} \label{jacmet}
d\tilde{s}^2 = 2\left[ E - V(q) \right]ds^2 \ .
\end{equation}
Now, Maupertuis' principle of least action 
can be stated as follows. The physical path of a mechanical system with conserved total energy $H=E$ is an extremal on the configuration space, i.e., 
\begin{equation} \label{maugeo}
\delta \int d\tilde{s} = \delta \int \sqrt{2\left[ E - V(q) \right]}ds = 0\ .
\end{equation}
 According to (\ref{jacmet}), the Jacobi line element can be expressed as
\be
d\tilde s ^2 = 
h _{\alpha\beta} d q^\alpha \otimes d q^\beta = 
2 \left[E-V(q)\right] g _{\alpha\beta} d q^\alpha \otimes d q^\beta \ ,
\ee
and the motion equations (\ref{geo1}) are completely equivalent to the geodesic equations
\be
\frac{d^2 q^\alpha}{d \tilde s^2} + \tilde \Gamma ^\alpha_{\ \beta\gamma} 
\frac{ d q ^\beta}{d\tilde s} \frac {d q ^\gamma}{d\tilde s} = 0 \ ,
\label{geo2}
\ee
where $\tilde  \Gamma ^\alpha_{\ \beta\gamma}$ is the Levita-Civita connection for the
Jacobi metric  $ h _{\alpha\beta}$, and $\tilde s$ can now be interpreted as an affine parameter along the geodesic that is related to the time parameter by
\be
\frac{d\tilde s}{dt} = 2 \left[E-V(q)\right] \ .
\ee

Assuming that the metric and the potential are smooth functions 
in the configuration space $M$, the physical trajectories must be contained in the open subspace 
\be
\Sigma=\{q^\alpha \in M : E>V(q)\}\ .
\ee
The boundary of this subspace is $\partial \Sigma =\{q^\alpha \in M : E=V(q)\}$. If a  trajectory reaches the boundary at a point 
$q_0^\alpha$, it must have zero velocity at that point. Consequently, there are allowed trajectories along the boundary. 
Physically, the boundary can be reached only if the potential has no critical points at $q_0^\alpha$. So, in principle, one could also include the boundary in the definition of the subspace $\Sigma$ of admissible physical trajectories. However,  the Jacobi metric vanishes on the boundary with the corresponding unpleasant geometric consequences. For this reason, in this work we will consider $\Sigma$ as an open subspace, and solutions of the geodesic equations (\ref{geo2}) are considered only in the interior of $\Sigma$. For a rigorous treatment of these variational principles, see \cite{arnold}. A generalization of Maupertuis' approach which includes the boundary of $\Sigma$ can be found in \cite{szy98}.

\section{The base space}
\label{sec:geo}
For the purposes of topological quantization, a fiber bundle must be constructed such that the base space contains all the information about the classical physical system. In this section,  we will find a base space with this characteristic for mechanical systems, and will analyze its geometric properties. 

The pair $(M,g)$ constitutes a smooth Riemannian manifold. 
The metric $g$ determines the kinetic energy by means of $T=(1/2)g(v,v)$, where $v$ is any vector in $T_q M, \ q\in M$. If we consider the motion of a particle, $v$ is the velocity vector of the particle with components $(\dot q ^1, \dot q ^2, ..., \dot q ^k)$. Choosing Cartesian-like orthonormal coordinates $\{x^\alpha\}$, the components of the metric reduce to $g(\partial_\alpha, \partial_\beta)=\delta_{\alpha \beta}$, with $\delta_{\alpha\beta}$  being the Euclidean metric, and the configuration manifold becomes flat. This result is valid independently of the potential $V(q)$. Consequently, the Riemannian manifold $(M,g)$ does not contain the information about the potential where the particle is moving. 
The phase space $T^*_q M$ is also not a suitable alternative, because the metric $g$ and the potential $V$, which contain all the information about the mechanical system, depend on the configuration variables only, and cannot be incorporated in a simple manner in the phase space. 

A different alternative would be to consider the set of all exact solutions of the geodesic-like equations (\ref{geo1}). This choice certainly contains all the information about the mechanical system, including the corresponding potential. However, exact solutions of this system of second order differential equations can be derived only for very particular potentials. In general, only approximate or numerical solutions are available. This makes the set of solutions difficult to handle. For the construction of a fiber bundle one needs as base space a set with, at least, the properties of a topological space. Even for the few cases in which the solutions of (\ref{geo1}) are known, it is not clear whether they build a topological space. 

In the preceding section we established via Maupertuis' principle a geometric set up for classical mechanics. We found that a mechanical system can be described by a Riemannian manifold $(\Sigma, h)$, where $\Sigma$ is an open submanifold of the configuration manifold $M$ and the physical trajectories are geodesics on $\Sigma$. The metric $h=2(E-V)g$ contains the potential explicitly in the conformal factor. In the special case of a free particle in Cartesian-like coordinates, we have that $h=2Eg = 2E\delta$, and $\Sigma$ becomes flat. For any nonzero value of the potential, the metric $h$ is conformally flat and contains all the physical information in the conformal factor. For more general situations in which $g$ is not flat, the geometry of $h$ can become more complicated, but still tractable as a Riemannian manifold. Consequently, the Riemannian manifold $(\Sigma, h)$ determines a classical mechanical configuration in the sense of \cite{netquev07}, and can be used as the base space for a fiber bundle.

To investigate the properties of this manifold, we consider an orientable orthonormal frame on $\Sigma$, $\{e_a\}$ where $a = 1, \ldots, k$, being $k$ the dimension of $\Sigma$. Two orthonormal frames with the same orientation are connected by an orthonormal transformation, $e'_a = (\Lambda^{-1})^{\ b}_{a} e_b$, where $\Lambda \in SO(k)$. This frame is orthonormal with respect to the metric $h$, if $h(e_a, e_b) = \delta_{ab}$. In addition, we introduce the dual basis $\{\theta^a\}$ which allows us to express the metric as
\begin{equation}  \label{orthomet}
h = h_{\alpha\beta}dq^\alpha \otimes dq^\beta = \delta_{ab} \,\theta^a \!\otimes \theta^b.
\end{equation}
In a torsion free manifold, we have that the connection one-form and the curvature two-form are given by Cartan's structure equations,
\begin{equation}  \label{cartan}
d\theta^a = - \omega^{a}_{\;\;b} \wedge \theta^b \ ,\quad
R^{a}_{\;\;b} = d\omega^{a}_{\;\,b} + \omega^{a}_{\;\;c} \wedge \omega^{c}_{\;\;b}\ ,
\end{equation} 
\noindent where $d$ is the exterior differentiation and 
$\omega^{a}_{\;\;b}:= \omega^{a}_{\ bc}\theta^c$ is a matrix valued one-form which is skew-symmetric in its two  indices $\omega_{ab} = -\omega_{ba}$. Connection and curvature are the main geometric objects that characterize the Riemannian manifold $\Sigma$. 
For later use, we calculate their components in the case of conformally flat metric, $h_{\alpha\beta}=2(E-V)\delta_{\alpha\beta}$. We choose the local dual frame as $\theta^a = e^{\varphi/2} \delta^a_{\ \alpha} d q^\alpha$, where $\varphi=\ln[2(E-V)]$. Then 
\begin{equation}   \label{oneformphi}
\omega_{ab} = e^{-\varphi/2}\frac{\partial \varphi}{\partial q^\beta}\delta^{\beta}_{\;\;[b}\delta_{a]c}\theta^c\ ,
\end{equation}
\begin{equation}  \label{twoformphi}
R_{ab} = e^{-\varphi}\left[ \delta^{\alpha}_{\;\;d}\delta^{\beta}_{\;\;[a}\delta_{b]c} \frac{\partial^2 \varphi}{\partial q^\alpha \partial q^\beta}  + \delta^{ef} \delta^{\alpha}_{\;\;[e}\delta_{a]c}\delta^{\beta}_{[b}\delta_{f]d} \frac{\partial \varphi}{\partial q^\alpha}\frac{\partial \varphi}{\partial q^\beta} \right] \theta^c \wedge \theta^d.
\end{equation} 
Finally, it is easy to prove that under a change of the   orthonormal dual frame
$\theta'=\Lambda \theta$, connection and curvature transform as
\begin{equation} \label{transform}
\omega' = \Lambda\omega\Lambda^{-1} + \Lambda d\Lambda^{-1} \qquad \text{and}  \qquad R' = \Lambda R \Lambda^{-1}\ ,
\end{equation}
where we have dropped the indices labeling the components in the orthonormal frame.

We will benefit from the introduction of an orthonormal frame for describing the geometry of the base space $\Sigma$ as will be clear in the next section. In section \ref{sec:hosc} we shall use some of these results to analyze some particular cases.
 
\section{The principal fiber bundle}
\label{sec:pfb}

In section \ref{sec:mau} we established that to any conservative mechanical system with $k$ degrees of freedom, there corresponds a $k$-dimensional Riemannian manifold $(\Sigma,h)$, whose metric contains a conformal factor with the potential function in it, and that 
distinct mechanical configurations are characterized by different potentials. When the metric is conformally flat, all the information is enclosed in the conformal factor. In the previous section we introduced an orthonormal frame on every point of $\Sigma$ to describe the geometry of the reduced configuration manifold. As it is known, most physical systems in classical mechanics are invariant under the Galilean group of transformations. The relevance of using orthonormal frames is that the Galilean group becomes locally reduced to its subgroup $SO(k)$. This is especially important for the construction of a fiber bundle, where the fiber represents the symmetries of the mechanical system. The main result of this section can be formulated as follows.

\textbf{Theorem:} A conservative mechanical system with $k$ degrees of freedom for which the Hamiltonian is a conserved quantity can be represented by a unique principal fiber bundle $P$ of dimension $\frac{1}{2}k(k+1)$, with the Riemannian manifold $(\Sigma, h)$ as the base space, where $h$ is the Jacobi metric and $\Sigma$ its domain in the configuration space, the group $SO(k)$ as the structure group (isomorphic to the standard fiber), and a connection with values in the Lie algebra $so(k)$. 

\noindent\textit{Proof:} The reconstruction theorem of differential geometry \cite{kn,naber} states that a fiber bundle is uniquely specified by the base space, the standard fiber, a structure group, and a family of transition functions, with values in the structure group, satisfying the cocycle condition. We will show that all these elements are present in our geometric construction of classical mechanics. The base space in this case is the $k$-dimensional reduced configuration manifold $\Sigma$ which, together with the Jacobi metric $h$, constitutes a smooth Riemannian manifold. The structure group is identified as the Lie group $SO(k)$ which is isomorphic to the standard fiber, in the case of a principal fiber bundle. It remains to show the existence of the transition functions. 

Let $\{U_i\}$ be a covering of $\Sigma$. Let $(U_i, \phi_i)$ be a chart in $\Sigma$, i.e.,
$U_i\subset \Sigma$ and $\phi_i: U_i \rightarrow V_i \subset \mathbb{R}^k$, where $U_i$ and
$V_i$ are open subsets and $\phi_i$ is a homeomorphism. This is the map that is commonly used in order to introduce a set of coordinates $q_i$ in the open subset $U_i$ of $\Sigma$. Let $(U_i , \tilde\phi _i)$ be the chart that allows us to introduce the dual frame $\theta_i$ in $U_i$, i.e., $\tilde\phi_i : U_i \rightarrow \Lambda^1(U_i,so(k))$. 
Notice that the index in $\theta_i$ denotes the set of dual frames in $U_i$ and does not refer to the components of the frame. If we consider the intersection 
$U_i \cap U_j \neq \emptyset$ with the corresponding dual frames $\theta_i$ and $\theta_j$,
and recall that all dual frames are related through a rotation, i.e., $\theta_i = \Lambda_{ij}\theta_j$ with $\Lambda_{ij}\in SO(k)$, it follows that  $\Lambda_{ij} = \tilde{\phi}_i \circ \tilde{\phi}_{j}^{-1}$ can be used as transition functions if they satisfy the cocycle condition. Indeed, in the case of a non empty triple intersection $U_i \cap U_j \cap U_k \neq \emptyset$, we also have the following maps that relate the frames on different open subsets $\Lambda_{kj} = \tilde{\phi}_k \circ \tilde{\phi}^{-1}_{j}$ and $\Lambda_{ik} = \tilde{\phi}_i \circ \tilde{\phi}^{-1}_{k}$. Then, it is obvious that the cocycle condition  (no summation over repeated indices)
\begin{equation}  \label{tripcond}
\Lambda_{ik}\Lambda_{kj} = \Lambda_{ij}
\end{equation}
is satisfied. It follows then from the reconstruction theorem that the fiber bundle $P$ 
with base space $\Sigma$, structure group $SO(k)$ is principal and unique. Since dim$(\Sigma)=k$ and dim$[SO(k)]=k(k-1)/2$, we obtain that dim$(P)=k(k+1)/2$, as stated in the above theorem.
 
To prove the existence of a connection on $P$, we only need to rephrase a well-known theorem (see, for instance, \cite{gockeler}, p. 150) stating that given an open covering $\{U_i\}$
of $\Sigma$, a family of local $so(k)$-valued one-forms, i.e., $\omega_i \in \Lambda^1(U_i, so(k))$ which fulfill the compatibility condition (no summation over repeated indices)
\be
\omega_i = \Lambda_{ij}  \omega_j \Lambda_{ij} ^{-1} 
+ \Lambda_{ij}  d \Lambda_{ij} ^{-1} \ , 
\label{comp}
\ee
where $\Lambda_{ij}:U_i \cap U_j \to SO(k)$ are elements of $SO(k)$, 
and a set of local sections $\sigma_i:U_i \to \pi^{-1}(U_i)$ satisfying $\sigma_i = \sigma_j \Lambda_{ij}$ on $U_i \cap U_j$, then there is a unique connection ${\bm {\omega}}$ on $P$ such that $\omega_i = \sigma^{*}_{i}\boldsymbol{\omega}$, where $\sigma^{*}_{i}$ is the pull-back induced by $\sigma_i$. 

Let us show that all the conditions of this theorem are satisfied. The base space $\Sigma$ is a Riemmanian manifold and therefore there must exist an open covering $\{U_i\}$. From the Jacobi metric $h$ we can construct 1-forms $\omega_i$ for each $U_i$, by using Cartan's first structure equation (\ref{cartan}), which take values in the algebra $so(k)$. If we apply a transformation of the dual frame of the form 
$\theta_i = \Lambda_{ij} \theta_j$, with $\Lambda_{ij} \in SO(k)$, then the transformation law (\ref{transform}) leads immediately to the compatibility condition (\ref{comp}).
It only remains to show the existence of local sections. 
As any other fiber bundle, our principal bundle $P$ is locally trivial, and  
accepts a local trivialization which can be defined
as $\Psi_i: \pi^{-1}(U_i) \to U_i\times SO(k)$. Then, a local canonical section 
can be defined as 
$\sigma_i: U_i \to \pi^{-1}(U_i)$, with  $\sigma_i(q) = \Psi_i^{-1}(q,e)$, where $e=\Lambda_{ii}(q)$ is the identity element of $SO(k)$ and 
$q\in U_i$. This canonical section satisfies the required condition $\sigma_i = \sigma_j \Lambda_{ij}$ on $U_i \cap U_j$, because all elements $\Lambda_{ij}$ are also  transition functions \cite{naber}. 
Then, we conclude that there exists a unique connection 
\mbox{\boldmath $\omega$} on $P$. This ends the proof of the theorem. 

\section{The topological spectrum}
\label{sec:hosc}
In the last section we showed that in any classical mechanical system there exists a triplet $(\Sigma, \omega, SO(k))$ which determines a unique principal fiber $P$ with a connection \mbox{\boldmath $\omega$}. It then seems plausible to use the invariant properties of $P$ to characterize each mechanical system. In particular, one can use the characteristic classes $C(P)$ which are topological invariants. Additionally, the integral $\int_\Sigma C(P)$ is also an invariant that can always be normalized such that  $\int_\Sigma C(P)=n$, where $n$ is an integer \cite{damas}. The characteristic classes for a principal fiber bundle with structure group $SO(k)$ are the Pontrjagin class $p(P)$ and the Euler class $e(P)$. Both can be written in terms of the curvature two-form $R$ of the base space $\Sigma$. 
The Pontrjagin class can obtained as the invariant polynomials in $R$ from 
\begin{equation}
\mathrm{Det}\left(It - \frac{{R}}{2\pi}\right) = \sum_{j=0}^{k} p_{k-j}({R})t^j \ ,
\end{equation}
and the Euler class is given as ($2m=k)$
\be
  e(P) = \frac{(-1)^m}{2^{2m}\pi^m m!} \epsilon_{i_1
   i_2 \dots i_{2m}}{R}^{i_1}_{\ i_2}\wedge
{R}^{i_3}_{\ i_4}\wedge\cdots \wedge
{R}^{i_{2m-1}}_{\ i_{2m}}  \ ,
\label{euler} 
\ee
and is different from zero only when $k$ is even. 

Since the explicit curvature for a mechanical systems is a function of the coordinates $q^\alpha$ and some parameters $\lambda_1,\ldots,\lambda_s$, which characterize the properties of the mechanical system, it is obvious that the integrals  $\int_\Sigma p(P)$ or $\int_\Sigma e(P)$ yield a relationship of the form $f(\lambda_1,\ldots,\lambda_s)=n$, where $n$ is an integer. This is the topological spectrum of the classical mechanical system. It implies a discretization of the physical parameters entering the mechanical system. In the following examples we calculate the topological spectra of some specific mechanical systems.

Consider the Lagrangian for two harmonic oscillators of mass $m$
\begin{equation} 
L = \frac{1}{2}m\left[ (\dot q ^1)^2 +  (\dot q ^2)^2 \right ]
- \frac{1}{2}\left[ k_1 (q ^1) ^2 + k_2( q ^2) ^2\right]\ ,
\end{equation} 
where $k_1$ and $k_2$ are constants. This system is conservative and corresponding Hamiltonian $H=E$ is a constant of motion. The two-dimensional metric $g=m\,{\rm diag} (1,1)$ is flat, and the Jacobi metric $h=2m(E-V) \, {\rm diag}(1,1)$ is conformally flat. Choosing the local dual frame as  $\theta^1 =\sqrt{2m(E-V)} dq^1$
and $\theta^2 =\sqrt{2m(E-V)} dq^2$, the symmetry of this classical configuration
reduces to transformations of the group $SO(2)$. The calculation of the curvature is straightforward, according to (\ref{twoformphi}), and the corresponding Euler class can be expressed as (a coma denotes partial
derivative)
\begin{equation}   
\label{euler1}
e(P) = -\frac{1}{2\pi}R_{12}  = 
\frac{1}{4\pi}(\varphi_{,11} + \varphi_{,22})dq^1 \wedge dq^2 \ ,
\end{equation}
\be
\varphi=\ln\left[2m\left( E- \frac{1}{2} k_1 (q ^1) ^2- \frac{1}{2} k_2( q ^2) ^2 \right)\right] \ .
\ee
The calculation is straightforward, but the resulting expression is 
quite cumbersome. For the sake of simplicity, we consider the special limiting case 
$k_2=0$, and let $k_1=k$ and $q^1=q$. Then
\begin{equation}  \label{euleronedim}
\int e(P) = -\frac{k }{4\pi}\int
\frac{E + \frac{1}{2}kq^2}{(E - \frac{1}{2}kq^2)^2}dq = n \ , 
\end{equation}
where we have chosen  as $\pi$ the constant resulting from the integration over $q^2$. 
Integrating over $q$ within the interval $[-q_0, q_0]$, we obtain the topological spectrum for the harmonic oscillator 
\begin{equation}   \label{quantcond}
\frac{kq_0}{kq_0^2 - 2E } = n \ , 
\end{equation}
where $q_0$ is a constant parameter related to the turning point of the oscillator. 
The specific choice \cite{netquev06}
\begin{equation}    \label{limit}
q_0 = \frac{1}{C} - \sqrt{\frac{1}{C^2} + \frac{2E}{k}} \ , \quad
 C = 2 \left(\frac{E}{\hbar\omega} - \frac{1}{2}\right)
\ , \quad \omega=\sqrt{\frac{k}{m}} \ ,
\end{equation}
transforms (\ref{quantcond}) into the canonical spectrum $
E= \hbar \omega(n+1/2)$. This shows that there exists a direct relationship between the topological spectrum and the canonical spectrum of the harmonic oscillator. 

Consider the Lagrangian for a particle of mass $m$ moving in a central field $V(r)$, i.e.,
\be 
L = \frac{1}{2} m (\dot r^2 + r^2\dot\vartheta^2) - V(r) \ .
\ee
The Hamiltonian of this system corresponds to the total energy
\be
E= \frac{1}{2} m \dot r^2 +  \frac{1}{2} \frac{l^2}{mr^2} + V(r) \ ,
\label{totale}
\ee
where $l$ is the angular momentum.  
The corresponding Jacobi metric is 
$h=2m(E-V)\,{\rm diag}(1,r^2)$, and the dual frame can be chosen as $\theta^r = \sqrt{2m(E-V)}\, dr$, and $\theta^\vartheta = \sqrt{2m(E-V)}\, r d\vartheta$. The only independent component of the curvature two-form is
\be
R_{r\vartheta} = \frac{1}{4m[E-V(r)] r}\frac{d}{dr}
\left( \frac{r}{E-V(r)}\frac{dV}{dr}\right) \, \theta^r \wedge \theta^\vartheta \ ,
\ee
so that the Euler class, $e(P) = -1/(2\pi) R_{r\vartheta}$,  can be integrated in general and yields 
\be
- \frac{1}{2}\left( \frac{r}{E-V(r)}\frac{dV}{dr}\right) \bigg|_{\Sigma_r} = n \ ,
\ee
where $\Sigma_r$ represents the domain of the coordinate $r$ in the base space $\Sigma$. The topological spectrum in this case can easily be calculated for any given potential $V(r)$. Consider, for instance, the Kepler potential $V(r)=-\alpha/r$ with $\alpha>0$. 
The analysis of the motion equations \cite{goldstein} 
shows that closed trajectories are allowed only if $E<0$. Then, the calculation of the topological spectrum 
\be
-\frac{\alpha}{2}\frac{1}{\alpha-|E|r} \bigg|_{\Sigma_r} = 
-\frac{\alpha}{2}\left( \frac{1}{\alpha-|E|r_+} - \frac{1}{\alpha-|E|r_-}\right) 
\ee
requires the values of the apsidal distances $r_-$ and $r_+$ which can be derived from the condition $\dot r =0$ and the expression for the total energy (\ref{totale}), i. e.,
\be
r_\pm = \frac{\alpha}{2|E|} \left( 1 \pm \sqrt{1-\frac{2|E|l^2}{m\alpha^2} } \right) \ .
\ee
The topological spectrum can finally be written as
\be
\frac { - \frac{2|E|l^2}{m\alpha^2} }{ \sqrt{1-\frac{2|E|l^2}{m\alpha^2} }} = \frac{1}{n} \  .
\ee
It is clear that this expression reproduces the behavior $|E|\sim 1/n^2$ of the canonical spectrum of a hydrogen-atom-like system.

\section{Conclusions}
\label{sec:con}
In this work we presented the first approach towards the topological quantization of classical mechanical systems. We found that Maupertuis's formalism of classical mechanics is appropriate for the development of a specific differential geometric structure which is the basis of topological quantization. In fact, to any given mechanical system we can associate a unique Jacobi metric which contains all the physical information of the system. The Jacobi metric determines a Riemannian manifold which we use  as the base space for a particular principal fiber bundle with a connection. The standard fiber is represented by the rotation group which contains all the symmetries of the mechanical system, when a local coframe is used to represent the corresponding Jacobi metric. This principal fiber bundle is shown to be unique and the corresponding connection on the bundle coincides with the connection of the Jacobi metric, when projected over the base space. For any given mechanical system we can calculate certain topological invariants which determine the topological spectrum of the mechanical configuration. In the case of a harmonic oscillator and the Kepler potential, we showed that the topological spectrum essentially reproduces the discrete behavior of the canonical spectrum which is usually obtained by applying the procedure of canonical quantization. 

Although these results are encouraging, topological quantization is not yet a complete procedure that could be used as an alternative to canonical quantization. Concepts like quantum states, wave functions, probabilities, etc. are essential in canonical quantization. These concepts are still under construction in topological quantization. Our goal is to develop a formalism by using all the geometric and topological information contained in the underlying principal fiber bundle. Preliminary results show that a certain family of local sections of the bundle can be used to define ``topological" quantum states. Probabilities must then be treated as probability distributions with certain Hamiltonian functions defined on those particular local sections. These questions are currently under investigation.

\section*{Acknowledgments}
This work was supported by Conacyt, Mexico, grant 48601.

\end{document}